\documentclass{PoS}
 \usepackage{epsfig}

\def\line{\hbox to \hsize}

 \newcommand\ts{\textstyle}

 \newcommand{\psib}{\overline{\psi}}

\title{Quark structure from the lattice Operator Product Expansion}

\ShortTitle{ %Short Title for header
       Quark structure from the lattice OPE}

 \author{W. Bietenholz{$^a$}, N. Cundy{$^b$},
 M. G\"ockeler{$^b$}, R. Horsley{$^c$}, 
 {H.~Perlt}{$^d$}, D.~Pleiter{$^e$}, \speaker{P.E.L.~Rakow}{$^f$},
 G.~Schierholz{$^{b,g}$}, A.~Schiller{$^d$}, T. Streuer{$^b$}
 and J.M. Zanotti{$^c$}\\

 \vspace{-7.2cm}
 {\line{\hfil {\rm  \large PoS(LAT2009)139}}}
 {\line{\hfil {\rm  \large DESY 09-208}}}
 {\line{\hfil {\rm  \large Liverpool LTH 855}}}
 {\line{\hfil {\rm  \large Edinburgh 2009/21}}}
 \vspace{6cm}

     {$^a$}
 Insituto de Ciencias Nucleares, Universidad Nacional
 Aut\'{o}noma de M\'{e}xico \\ \quad
    A.P.\ 70-543, C.P.\ 04510 Distrito Federal, M{e}xico \\
          \llap{$^b$}Institut f\"ur Theoretische Physik,
  Universität Regensburg,
   93040 Regensburg, Germany\\
    \llap{$^c$}
   School of Physics, University of Edinburgh, Edinburgh EH9
 3JZ, United Kingdom \\
          \llap{$^d$}Institut für Theoretische Physik,
  Universit\"at Leipzig,  04109 Leipzig, Germany\\
  \llap{$^e$} John von Neumann Institut f\"{u}r Computing 
 NIC/DESY Zeuthen, 
 15738 Zeuthen, Germany \\
          \llap{$^f$} Theoretical Physics Division,
  Department of Mathematical Sciences, 
 University of Liverpool,Liverpool L69 3BX, UK\\
          \llap{$^g$} Deutsches Elektron-Synchrotron DESY,
   22603 Hamburg, Germany \\
       E-mail: \email{rakow@amtp.liv.ac.uk}\\

 {\large \bf QCDSF Collaboration}
 }

\abstract{  We have reported elsewhere in this conference on our
 continuing project to determine non-perturbative Wilson
 coefficients on the lattice, as a step towards a completely
 non-perturbative determination of the nucleon structure.
 In this talk we discuss how
 these Wilson coefficients can be used to extract Nachtmann
 moments of structure functions, using the case of off-shell
 Landau-gauge quarks as a first simple example.
     This work is done using overlap fermions, because their
 improved chiral properties reduce the difficulties due to
 operator mixing.}

\FullConference{The XXVII International Symposium on
 Lattice Field Theory - LAT2009\\
		 July 26-31 2009\\
		 Peking University, Beijing, China}

\begin{document}

\section{Introduction}

     W. Bietenholz has explained our procedure for extracting
 Wilson coefficients  from lattice measurements 
 in his proceedings~\cite{WBtalk}. In this report we will 
 show how these coefficients can be used to reconstruct Compton scattering
 amplitudes, and to extract their Nachtmann moments. We start with a 
 toy example, looking at the Nachtmann moments of off-shell quarks 
 in Landau-gauge background fields. 

     The calculations we report here are done with overlap fermions
 on a quenched background. Overlap fermions were chosen for their
 superior chiral symmetry properties. The negative mass parameter 
 is $\rho = 1.4$. We used a $24^3 \times 48$ lattice, with a lattice
 spacing $a = 0.095$~fm. The results shown here all have bare 
 mass $ a m_q = 0.028 $. We have always taken the scattering momentum 
 $q$ along a lattice diagonal, $q \propto ( 1, 1, 1, 1 ) $ for
 maximum lattice symmetry. We employ three different values for the 
 magnitude of $a^2 q^2$, so that we can begin to investigate scaling
 in $q^2$. All the Green's functions
 have been $O(a)$ improved using the prescription of~\cite{Capitani}.
 The electromagnetic current
  $J_\mu(x)$ is represented by the local current
  $\psib(x) \gamma_\mu \psi(x)$, with overlap $O(a)$ improvement.

 \section{Operator Product Expansion} 

  We express the electromagnetic scattering tensor $W_{\mu \nu}$ as
 a sum over local operators (the Operator Product Expansion or OPE).
 In each term we have a separation 
 of scales, all dependence on the quark momentum is in the 
 matrix element
 $\langle \psi(p) | J_\mu(q) J^\dag_\nu(q) | \psi(p) \rangle$,
 all dependence on the photon scale $q$ is in the Wilson coefficient
 $C_{\mu \nu}^m(q)$. At present we are only considering flavour non-singlet
 processes, so we do not include any purely gluonic operators in the sum, 
 \begin{equation}
 W_{\mu \nu}(p,q)
  \equiv \langle \psi(p) | J_\mu(q) J^\dag_\nu(q) | \psi(p) \rangle 
 = \sum_m C_{\mu \nu}^m(q) \langle \psi(p) | {\cal O}^m| \psi(p) \rangle
 \;.
 \end{equation} 
 We include quark bilinear operators ${\cal O}^m$ with up to 3 covariant
 derivatives in this sum. When 
 all possible Dirac structures are taken account of, there are
 potentially 1360 different operators, and 1360 Wilson coefficients
 $C^m_{\mu \nu}$, 
 in the sum. We reduce this number by exploiting lattice symmetries. 
 We choose $q$ along a lattice diagonal, i.e.~$q \propto (1,1,1,1)$. 
 With this choice there are only 67 independent Wilson coefficients
 in the expansion of a diagonal element of $W$, such as
 $W_{44}$~\cite{Lat07}. 

  $ W_{\mu \nu}(p,q)$  is a fairly complicated object, it depends
 on $p, q, \mu, \nu$ and on the Dirac indices of the incoming
 and outgoing quark. 

  We simplify by just looking at unpolarised quarks (later, we plan to
 analyse spin-dependent quantities too). Taking the trace
 \begin{equation}
  T_{\mu \nu}(p,q)
 \equiv \frac{1}{4}  {\rm Tr}\left\{ S^{-1}(p) W_{\mu \nu}(p,q) \right\}\,
   ( p^2 + m^2 ) \;, 
  \end{equation}
 where $S^{-1}$ is the inverse quark propagator,
  removes the Dirac-index structure. The propagator cancels all 
  $Z_\psi$ factors, the only renormalisation we need is a factor
 of $ Z_V^2 $ to correct for using local currents for $J_\mu$.

   If we consider unpolarised quarks there are 4 tensor structures
 which can occur in the scattering tensor:
 \begin{equation}
 T_{\mu \nu} 
 = \delta_{\mu \nu} W_1 + p_\mu p_\nu W_2
 + (p_\mu q_\nu + q_\mu p_\nu ) W_4 + q_\mu q_\nu W_5 \;. 
  \end{equation}
  ($W_3$ and $W_6$ are reserved for structures possible in
 neutrino scattering and the polarised target case.)
  The $W_i$ form factors can only depend on invariants
  $q^2,\; p \cdot q,\; p^2 $. When we consider scattering on
 physical hadrons, we can use electromagnetic gauge invariance to
 reduce the expansion from four terms to two, namely $F_1$ and $F_2$. 
 When however we consider an off-shell quark this argument
 no longer applies, and all four structures are independent. 

   As a first simple case we consider the polarisation trace 
 of $T_{\mu \nu}$, 
  \begin{equation}
 T_{\mu \mu} 
 = 4 W_1 + p^2 W_2 + 2 p\cdot q W_4 + q^2 W_5 \;. 
  \end{equation}
 Advantages of this choice are that averaging over the
 direction of $J_\mu$
 (the photon polarisation) simplifies the rotation group theory
 considerably, and that this quantity only involves diagonal elements
 of $W$, which we have analysed more completely. (We have gathered data on
 off-diagonal components, $\mu \ne \nu$, but this has not yet been 
 fully analysed.) Summing over all polarisations also increases the
 symmetry, there are only 22 independent Wilson coefficients in the
 OPE  of $T_{\mu \mu}$, which is a considerable reduction compared 
 to the 67 coefficients needed for a single diagonal component
 such as $T_{44}$. 

    In~\cite{Nachtmann} Nachtmann proposed some quantities 
 (the Nachtmann moments) with 
 particularly simple Operator Product Expansions. 
 Nachtmann considered scattering from on-shell targets, 
 but we are interested
 in off-shell targets, so we have to generalise the formulae
 in~\cite{Nachtmann}. 
 The Nachtmann moments, $\mu_n,$
 are defined by splitting $T$ up into components of
 definite spin, $n$, 
 \begin{equation}
 T_{\mu \mu}( p\cdot q, q^2, p^2) = 2 \sum_{n {\rm \ even}}
  \left(\frac{p^2}{q^2} \right)^{\frac{n}{2}} 
 U_n(\cos \theta) \mu_n(q^2, p^2) 
 \label{Ndef}
 \end{equation}
 where $\theta$ is the angle between $p$ and $q$,
 \begin{equation}
 p \cdot q = |p|\; |q| \; \cos \theta
 \end{equation}
 and $U_n$ is a Chebyshev polynomial of the second type,~\cite{AbSteg}.
 The $U_n$ are the 4-dimensional equivalent of the familiar 3-dimensional
 spherical harmonics. 
  We can use orthogonality of the Chebyshev polynomials to project
 out single   Nachtmann moments from~(\ref{Ndef}),
 \begin{equation}
 \int_0^\pi \frac{d \theta}{\pi}\; \sin^2 \theta \; U_n(\cos \theta) \;
 T_{\mu \mu}(  p\cdot q, q^2, p^2)
 = \left( \frac{p^2}{q^2} \right)^{\frac{n}{2}}
 \mu_n (q^2, p^2) \;.
 \label{euclid}
 \end{equation} 
 The Euclidean integral~(\ref{euclid}) involves real values of $p\cdot q$
 in the range $-|p|\,|q| \le p\cdot q \le |p|\,|q|$.

 \begin{figure}[htb]
 \begin{center}
 \epsfig{file=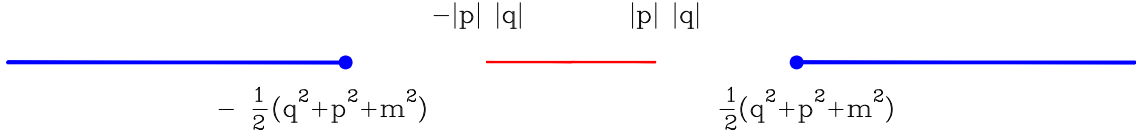,angle=270,width=15cm}
 \vspace*{2cm}
 \caption{{\it The complex $p\cdot q$ plane, for $p^2, q^2$ both
 positive (space-like). The Euclidean integral~(2.7) %(\ref{euclid})
 for the Nachtmann
 moment $\mu_n$ runs between $p\cdot q = -|p|\,|q|$ and
 $p\cdot q = +|p|\,|q|$, the Minkowski
 integral~(2.9)   %(\ref[Mink1])
 runs along the cut from the threshold at $p\cdot q =
 \frac{1}{2}(q^2 + p^2 + m^2) $ to $\infty$.}
  \label{pqplane} 
  }
 \end{center}
 \end{figure}

 Following~\cite{Nachtmann} we can relate the integral to Minkowski physics
 by allowing $p\cdot q$ to become complex, while keeping both $p^2$ and
 $ q^2$ real and positive, see Fig.\ref{pqplane}. Ignoring the possible
 complications due to confinement, we expect the amplitude $T$ to have
 branch-points at the thresholds for producing on-shell quarks,
 at $p \cdot q = \pm \frac{1}{2}(q^2 + p^2 + m^2)$, with cuts reaching out
 to infinity. We define the discontinuity across these cuts by
 \begin{equation}
 2 \pi i D( p\cdot q, q^2, p^2, m) \equiv
 T_{\mu \mu}( p\cdot q + i \varepsilon, q^2, p^2)
 - T_{\mu \mu}( p\cdot q - i \varepsilon, q^2, p^2 ) \;. 
 \label{discon}
 \end{equation}
  For colour singlet hadrons this discontinuity is a physically
 measurable total cross-section, but of course the cross-section 
 for deep inelastic scattering on an off-shell quark target
 is something we can only measure as a {\it Gedankenexperiment}. 
   Assuming that $T$ is an analytic
 function of the complex variable
 $p\cdot q$ we can write down dispersion relations
 which give the result of the Euclidean integral~(\ref{euclid}) as
 an integral involving the discontinuity (valid for even $n$, $n \ge 2$).
 \begin{equation}
 \label{Mink1}
 \!\! \mu_{n}(q^2, p^2)
 = 2 \int_\Theta^\infty \! d (p\!\cdot\!q)\,
  \frac{ D( p\cdot q, q^2, p^2)\; (q^2)^{n} }
 { \left(p \cdot q + \sqrt{ (p \cdot q)^2 - p^2 q^2}\right)^{n+1} }
 \end{equation}
 where $\Theta$ is the threshold for the production of on-shell
 particles,
 \begin{equation}
 \Theta \equiv {\ts \frac{1}{2} } ( q^2 + p^2 + m^2) \;.
 \end{equation}
  In general the Minkowski integral~(\ref{Mink1}) is complicated, 
 but in  the Bjorken limit, when $(p \cdot q)^2 \gg p^2 q^2$, we can
 simplify~(\ref{Mink1}) by changing to the integration variable
 \begin{equation}
 x \equiv \frac{q^2}{2 p \cdot q} \;.
 \end{equation}
 Eq.~(\ref{Mink1}) becomes
 \begin{equation}
 \mu_{n}(q^2, p^2)  =  \int_0^{\frac{q^2}{2 \Theta} } dx \;
 \frac{  x^{n-1} D(p\cdot q, q^2, p^2)}
 { \left[ \frac{1}{2} + \frac{1}{2} \sqrt{ 1 - 4 \frac{p^2}{q^2} x^2}
 \;\; \right]^{n+1}} \  \to 
   \int_0^1 dx \; x^{n-1} D(p\cdot q, q^2, p^2) \label{xmom}
 \end{equation}
 and we see that the Nachtmann moment tends to a simple $x$ moment.

    In the rest of this report we concentrate on the smallest
 interesting spin, $n=2$. From~(\ref{xmom})
 we see that at large $q^2$, $\mu_2$
 corresponds to $\langle x \rangle$. 

    In our discussion of Nachtmann moments we have assumed 
 full rotation symmetry. However, lattice operators are classified 
 under the hypercubic group,  and will normally be mixtures of
 representations of the full Euclidean rotation group. The 22
 operators present in the expansion of $W_{\mu \mu}$ fall into
 three hypercubic classes

 \begin{center}
 \begin{tabular}{rrrr}
 spin 0, &         & spin 4, \ + higher & \qquad 8 operators\\
         & spin 2, & spin 4, \ + higher & \qquad 13 operators\\
         &         & spin 4, \ + higher & \qquad 1 operator
 \end{tabular} 
 \end{center}
 For our first look at spin 2, we simply keep all the operators in
 the middle group, and discard the others. 
  The leading spin 2 operator (for $q  \propto (1,1,1,1)$ ) 
 has the form
 \begin{equation} 
 \sum_{\mu \ne \nu}  \psib ( \gamma_\mu D_\nu + \gamma_\nu D_\mu)\psi
 \label{v2a}
 \end{equation} 
 i.e.~it is a symmetric, off-diagonal tensor. 

    We now reconstruct a projected spin 2 amplitude, 
 $[T_{\mu \mu}]_{spin\ 2}$, by taking the product of the 
 matrix elements of the 13 spin 2 operators with their
 Wilson coefficients, as determined in~\cite{WBtalk}. 

 \begin{figure}[htb]
 \begin{center}
 \epsfig{file=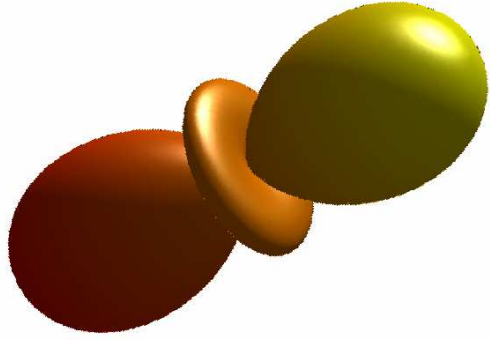,width=8cm}
 \end{center}
 \caption{\it The 4-dimensional spherical harmonic
 $ p^2 q^2 U_2(\cos \theta) = 4 (p\cdot q)^2 - p^2 q^2$.
 The  main lobes point along the directions $\pm q$. There are nodes
 when $p$ is $60^{\rm o}$ and $120^{\rm o}$ from $q$; the equatorial    
 `donut' has the opposite sign to the main lobes. 
 \label{harmonic}}
 \end{figure}

 Because we have averaged over quark spin and photon polarisations,
 the only direction left in our problem is $q \propto (1,1,1,1)$ 
 and the only spin-2 4-d spherical harmonic that contributes is 
 \begin{equation} 
  p^2 q^2 U_2(\cos \theta) = 4 (p\cdot q)^2 - p^2 q^2  \;, 
 \end{equation}
 which is illustrated in Fig.{\ref{harmonic}}. 
 When we put in the fact that $q \propto (1,1,1,1)$ 
 \begin{equation} 
  4 (p\cdot q)^2 - p^2 q^2  
 \to  2 \left( p_1 p_2 + p_1 p_3 + \cdots + p_3 p_4 \right) q^2 \;. 
 \label{harmon}
 \end{equation} 
 Spin 2 operators, such as~(\ref{v2a}), should have an 
 expectation value proportional to~(\ref{harmon}). 

   We can find the $\mu_2$ for the quark from~(\ref{euclid}), 
 \begin{equation}
 q^2 \left[ T_{\mu \mu} \right]_{spin 2} = 2 \;
 \left( \sum_{\mu < \nu } p_\mu p_\nu \right) \; \mu_2 \;, 
 \label{xdef} 
  \end{equation}
 so if we plot $ q^2 \left[ T_{\mu \mu} \right]_{spin 2} $
 against $2 \; \left( \sum_{\mu < \nu } p_\mu p_\nu \right) $ 
 we should see a straight line passing through the origin, 
 with a slope equal to $\mu_2$. 

    We have data for three different $q$ values, and from 15 to
 32 $p$ values (depending on $q$),
 chosen to give a good coverage of directions, so we
 can see whether the projected amplitude really follows the spherical
 harmonic, and find $\mu_2$ for a quark.

 \section{Results} 

  \begin{figure}[htb]
  \begin{center}
 \epsfig{file=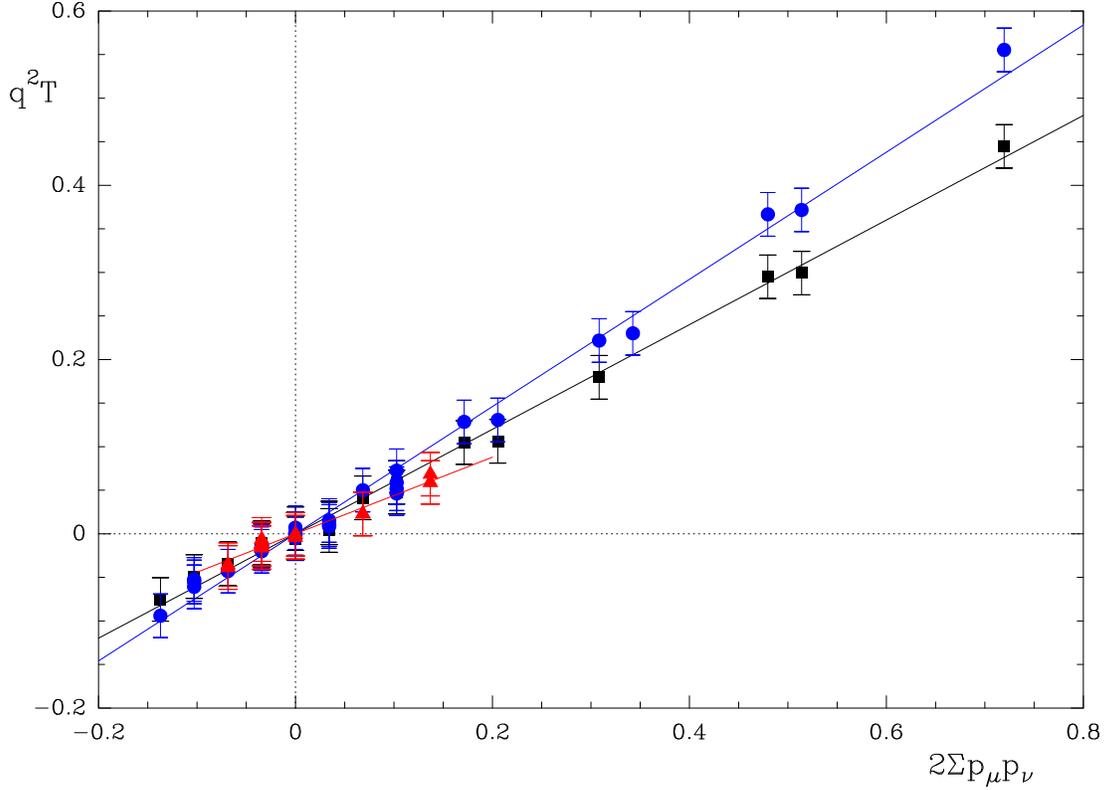, angle=270, width=14.5cm}
 \end{center}
 \caption{\it The trace  $ q^2 \left[ T_{\mu \mu} \right]_{spin 2} $
 plotted 
 against $2 \; \left( \sum_{\mu < \nu } p_\mu p_\nu \right) $,
 in lattice units. 
 These points should fall on a straight line through the origin,
 with a slope 
 proportional to $\mu_2$, the Nachtmann moment corresponding 
 to $\langle x \rangle$, see eq.(2.16). The black squares
 are data from $q^2 = \pi^2/(2.25 a^2) \approx 19$ GeV$\,^2$,
 blue circles, 
  $ q^2 = \pi^2/(4 a^2) \approx 10.6$ GeV$\,^2$, red triangles 
 $ q^2 = \pi^2/(9 a^2)\approx 4.7$ GeV$\,^2$. 
 \label{fullq1q2q3}} 
 \end{figure} 

 In Fig.\ref{fullq1q2q3} we show the spin-projected Compton amplitude
 plotted against $2 \; \left( \sum_{\mu < \nu } p_\mu p_\nu \right) $. 
 The quantity  $2 \; \left( \sum_{\mu < \nu } p_\mu p_\nu \right) $
 is positive for momenta in the main lobe of the harmonic in
 Fig.\ref{harmonic}, zero for momenta in the node, and negative for
 momenta in the equatorial ring. 

  At each $q^2$ value the amplitudes follow a straight line rather 
 closely. 
 Several quarks have momenta at  exactly $60^{\rm o}$ angle to $q$, 
 these fall on the node of the spherical harmonic, (zero on the horizontal
 axis), and they 
 have amplitudes very close to 0, as predicted. 
 Data from all three $q$ values are plotted, we see that they 
 scale fairly well. $q^2$ changes by a factor of 4, from 
 4.7~GeV$^2$ to 19~GeV$^2$. 

    Rough values for the Nachtmann moment $\mu_2$ (which is a 
 measure of $\langle x \rangle$ at a scale $\sim q^2$) are

 \begin{center}
 \begin{tabular}{cc}
 $q^2 = 4.7$ GeV$^2$ \qquad  & \qquad $\mu_2 = 0.44(9) $ \\ 
 $q^2 = 10.6$ GeV$^2$ \qquad & \qquad $\mu_2 = 0.73(5) $ \\ 
 $q^2 = 19$ GeV$^2$ \qquad & \qquad $\mu_2 = 0.60(5) $  
 \end{tabular}
 \end{center}

 \noindent These values probably include fairly large lattice artefacts
 $\propto a^2 q^2$. We will try to correct for these by looking 
 at tree-level lattice artefacts, which might lead to significant
 numerical changes. 

 \section{Conclusions and Prospects} 

   We have seen how we can project out spin components 
 for Nachtmann moment operator expansions. In the channel 
 we looked at (spin 2), simply filtering on the lattice 
 symmetry seems to work fairly well, we don't see any 
 sign of spin 4 contamination distorting the straight-line
 behaviour of Fig.\ref{fullq1q2q3}. 

   We don't see any strong dependence of $\mu_2$
 on the quark virtuality $p^2$, this may be a little unexpected. 

  There are several more things we could do. Here we have concentrated 
 on the Nachtmann moment coming from the polarisation trace
 $W_{\mu \mu}$, we should also look at the moments for the 
 other components of  $W_{\mu \nu}$. Particularly interesting 
 in the context of off-shell quarks would be to look at the 
 OPE for $q_\mu W_{\mu \nu} q_\nu $. This should give a 
 combination of operators which can be non-zero for off-shell 
 quarks, but which should vanish on-shell. The three-point
 functions for this combination of operators should show 
 contact terms, but no plateau or exponentially decaying terms. 

  The antisymmetric parts of the scattering tensor, 
 $W_{\mu \nu} - W_{\nu \mu} $ contain the information needed
 to investigate spin-dependent structure functions. We have
 collected the data needed for this calculation. 

     From looking at our problem in tree level we see that 
 $O(a^2 q^2)$ artifacts may be important in some channels. 
 We want to investigate these artifacts further, and see 
 whether we can use tree-level results to reduce their
 severity. 

 \section*{Acknowledgements}

 This work was supported by the Deutsche Forschungsgemeinschaft (DFG) 
 through project FOR 465 ``Forschergruppe Gitter-Hadron-Ph\"anomenologie". 
 The computation for this project was carried out on the computers of 
 the ``Norddeutscher Verbund f\"ur Hoch- und H\"ochstleistunsrechnen'', 
 (HLRN).


\begin{thebibliography}{99}

\bibitem{WBtalk} W. Bietenholz, proceedings of this conference. 
 [arXiv:0910.2437 [hep-lat]]


 \bibitem{Capitani}
  S.~Capitani, M.~G\"ockeler, R.~Horsley, P.~E.~L.~Rakow and G.~Schierholz,
  %``Operator improvement for Ginsparg-Wilson fermions,''
  Phys.\ Lett.\  B {\bf 468} (1999) 150
  [arXiv:hep-lat/9908029].
  %%CITATION = PHLTA,B468,150;%%

 \bibitem{Lat07}
  W.~Bietenholz {\it et al.},
  %``The operator product expansion on the lattice,''
  PoS {\bf LAT2007} (2007) 159
  [arXiv:0712.3772 [hep-lat]].
  %%CITATION = POSCI,LAT2007,159;%%

\bibitem{Nachtmann} O.~Nachtmann,
  %``Positivity constraints for anomalous dimensions,''
  Nucl.\ Phys.\  B {\bf 63} (1973) 237.
  %%CITATION = NUPHA,B63,237;%%

  \bibitem{AbSteg}
 Abramowitz and Stegun, ``Handbook of Mathematical Functions'', (Dover)
 1972.  Chapter 22.


\end{thebibliography}
\end{document}